# Exploratory Data Analysis (EDA) for Banking and Finance: Unveiling Insights and Patterns

Ankur Agarwal \*, Shashi Prabha and Raghav Yadav

Department of Computer Science Information Technology

, Sam Higginbottom University of Agriculture, Technology and Sciences,

Prayagraj, 21107, Uttar Pradesh, India.

\*Corresponding author(s). E-mail(s): 22phcoa101@shiats.edu.in;

Contributing authors: dr.shashiprabha@shiats.edu.in;

raghav.yadav@shiats.edu.in;

## Abstract

This paper explores the application of Exploratory Data Analytics (EDA) in the banking and finance domain, specifically focusing on analyzing credit card usage and customer churning. The report presents a comprehensive step-by-step analysis of credit card data using various EDA techniques, including descriptive statistics, data visualization, and correlation analysis. The analysis of credit card usage delves into transaction patterns, shedding light on the frequency, volume, and types of transactions performed by customers. Additionally, the paper examines credit limits and their distribution among cardholders, identifying potential areas for optimization. Furthermore, the report investigates the usage of credit cards across different merchant categories, providing insights into consumer spending behavior and preferences. Moreover, the analysis considers the influence of demographic factors, such as age, gender, and income, on credit card usage patterns. In addition to credit card usage analysis, this report delves into the critical issue of customer churning in the banking sector. Churn rate analysis uncovers the proportion of customers who discontinue with the bank within a given period. By identifying the churn rate, financial institutions can assess customer retention performance and set benchmarks for improvement. The report further investigates the factors driving customer churn, including customer demographics, transaction history, and customer satisfaction levels. These insights enable banking and finance professionals to make data-driven decisions, formulate targeted marketing strategies, and design effective customer retention initiatives. The findings from this analysis contribute to the advancement of the banking and finance sector's ability to enhance customer satisfaction, optimize credit card services, and ultimately drive profitability

**Keywords:** Exploratory Data Analytics (EDA), customer churning, descriptive statistics, data visualization, correlation analysis, data-driven decision-making, customer retention.

## 1. Introduction

In today's data-driven world, organizations and researchers are inundated with vast amounts of data. Data analysis is the process of inspecting, cleaning, transforming, and modelling data to uncover useful information, patterns, and insights. It involves applying various statistical, mathematical, and computational techniques to understand the data and draw meaningful conclusions [1].

Data analysis is crucial in banking and finance as it helps institutions make data-driven decisions, manage risks, detect fraud, assess creditworthiness, optimize operations, and gain insights into customer behavior. It enables banks and financial institutions to leverage the vast amount of data they collect to improve efficiency, identify opportunities, and enhance overall performance. Data Analysis can be further divided into two types i.e. Classical Data Analysis and Exploratory Data Analysis [2].

Classical data analysis (CDA) refers to the traditional approach of analyzing data using well-established statistical techniques and models. It follows a structured and hypothesis-driven approach, where specific hypotheses are formulated and tested using statistical tests and models. Classical data analysis involves making formal inferences, drawing conclusions, and making decisions based on the results of these statistical tests. It emphasizes hypothesis testing, parametric assumptions, and statistical significance.

Exploratory Data Analysis (EDA) is an approach to analyzing data that focuses on exploring and understanding the data without preconceived notions or specific hypotheses. It involves visual exploration, summary statistics, data transformations, and other techniques to gain insights and discover patterns in the data.

Exploratory Data Analysis (EDA) is more flexible, intuitive, and iterative compared to classical data analysis. Its primary goal is to summarize and visualize the data, detect patterns, identify relationships, and generate hypotheses for further investigation.

Exploratory Data Analysis (EDA) is particularly useful when dealing with messy, incomplete, or unstructured data. Fig. 1 shows the difference between Classical Data Analysis and Exploratory Data Analysis [3].

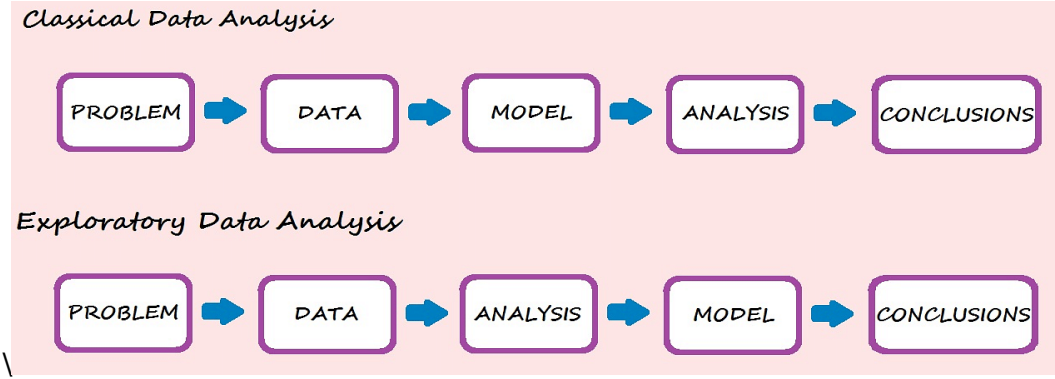

Figure 1: Difference between Classical Data Analysis and Exploratory Data Analysis

While classical data analysis has its own merits and is suitable for hypothesis testing and formal inference, EDA provides a more exploratory and intuitive approach to understanding the data, generating insights, and formulating research questions [4]. this paper focuses on Exploratory Data Analysis and its applications in Banking and Finance. Section 2 of the paper reviews prior research and introduces the fundamental concepts and objectives of Exploratory Data Analysis (EDA) in the context of banking and finance, setting the stage for deeper exploration. Section 3 discusses the various techniques used in EDA, including data cleaning, preprocessing, and statistical methods that are essential for analyzing complex datasets effectively. In Section 4, the paper explores how EDA is applied to specific banking functions, illustrating its role in enhancing operational effectiveness and strategic decision-making. Section 5 focuses on the application of EDA to analyze customer churn, detailing how data analysis can identify patterns and factors influencing customer retention. Section 6 addresses the challenges and critical considerations in implementing EDA, such as data quality, privacy, and compliance with industry regulations, underscoring the complexities of deploying EDA in a regulated environment. Finally, Section 7 provides concluding remarks that summarize the insights derived from the application of EDA and highlights its value in fostering data-driven strategies within the banking sector.

## 2. Related Work

Numerous research papers exist on the use of EDA in banking and finance, focusing on gaining insights, identifying patterns, and understanding the underlying dynamics of financial data to support informed decision-making [5]. Exploratory Data Analysis (EDA) serves as a cornerstone in the banking and finance sectors, playing a pivotal role in shaping accurate forecasts, assessing risks, and detecting anomalies with robustness. John W. Tukey's groundbreaking work, published in 1977, laid the foundational bedrock for EDA by introducing concepts and techniques that advocate for an iterative approach of plotting, modeling, and validating, ultimately revealing hidden data structures and outliers [6]. This iterative methodology, pioneered by Tukey, forms the backbone upon which subsequent advancements in EDA have been built. Expanding upon Tukey's seminal contributions, William S. Cleveland's "Visualizing Data" (1993) underscored the indispensable nature of visual tools in uncovering patterns, particularly crucial for dissecting the intricate distributions and anomalies inherent in financial datasets [7]. Cleveland's emphasis on visualization not only enhanced the interpretability of data but also facilitated a deeper understanding of complex financial phenomena, thereby empowering analysts to make informed decisions.

In "Data Science for Business," Foster Provost and Tom Fawcett (2013) further expound on the practical applications of EDA within business contexts, particularly within banking and finance, elucidating how a nuanced comprehension of data can catalyze decision-making processes [8]. By bridging the gap between theoretical concepts and real-world applications, Provost and Fawcett's work underscores the transformative potential of EDA in driving strategic initiatives and fostering innovation within the financial landscape.

Moreover, EDA plays a pivotal role in unraveling customer behavior and preferences within the banking and finance sectors, thus facilitating targeted marketing strategies aimed at enhancing customer satisfaction and fostering long-term loyalty. A plethora of research endeavors delve into various methodologies and insights in this domain, exemplified by Gupta et al.'s exploration of machine learning techniques for customer segmentation [9]. Additionally, Wibowo et al. leverage predictive analytics to discern consumer behavior patterns, thereby optimizing marketing strategies and bolstering retention efforts [10]. Ahmed et al.'s application of data mining methods further augments these efforts by enabling the prediction of customer behaviors, thereby facilitating more effective marketing campaigns and service enhancements [11]. K¨onigstorfer et. al in their research work explored applications of Artificial Intelligence in commercial banks and investigated behavioral finance to understand retail banking customers' decisions [12].

Abdolvand focused on customer lifetime value modeling to refine marketing strategies and enhance profitability [13]. These studies provide a comprehensive view of how EDA and associated analytical methods can be leveraged to improve understanding and engagement with banking customers.

Exploratory Data Analysis (EDA) significantly enhances decision-making processes in banking and finance, influencing areas such as product development, pricing strategies, portfolio management, investment decisions, and resource allocation. Many reserachers delve into how EDA techniques improve financial decision support systems, aiding in more nuanced investment and financial product decisions [14]. Tatsat et al in their recent work discussed the benefits of EDA in optimizing portfolio strategies and managing investment risks [15].

Furthermore, EDA extends its influence into diverse realms of financial decisionmaking, permeating areas such as product development, pricing strategies, portfolio management, investment decisions, and resource allocation. Hasan et al.'s exploration of big data analytics, including EDA, sheds light on how these methodologies contribute to the development of competitive pricing strategies and the creation of innovative product offerings [16].

In addition to strategic decision-making, EDA also serves as a catalyst for technological innovation within the financial landscape, as exemplified by Levenberg et al.'s exploration of text mining and sentiment analysis for predicting economic indicators [[17]. Similarly, Kim et al.'s proposition of a financial time series forecasting model utilizing Support Vector Machines underscores the efficacy of advanced machine learning techniques in augmenting traditional forecasting methodologies [18]. A comprehensive review of data mining techniques used in financial markets is provided by Kumar et al. [19]. Masini extends machine learning applications beyond traditional time series analysis in financial predictions [20]. Collectively, many of these research endeavors underscore the multifaceted nature of EDA's impact on financial markets, transcending traditional boundaries to drive innovation and foster informed decision-making [21] [22].

In conclusion, the literature review presented herein underscores the indispensable role of Exploratory Data Analysis (EDA) in revolutionizing decision-making processes within the banking and finance sectors. From its foundational principles laid down by visionaries like John W. Tukey and William S. Cleveland to its contemporary applications elucidated by scholars such as Foster Provost, Tom Fawcett, and numerous others, EDA has emerged as a cornerstone for understanding, interpreting, and leveraging complex financial data. Through its iterative approach, emphasis on visualization, and integration with advanced analytical techniques, EDA not only facilitates the dissection of customer behavior and preferences but also drives strategic initiatives ranging from product development to portfolio management. Moreover, the synthesis of diverse methodologies, from machine learning to sentiment analysis, underscores EDA's transformative potential in shaping the future landscape of banking and finance. As the financial ecosystem continues to evolve, EDA stands as a beacon of innovation and insight, empowering stakeholders to navigate the complexities of an ever-changing market with confidence and precision.

## 3. Key Techniques in Exploratory Data Analytics

Exploratory Data Analysis (EDA) encompasses several key techniques that aid in uncovering insights and patterns within data. Important techniques that are employed during EDA to achieve its objectives are indicated in figure 2:

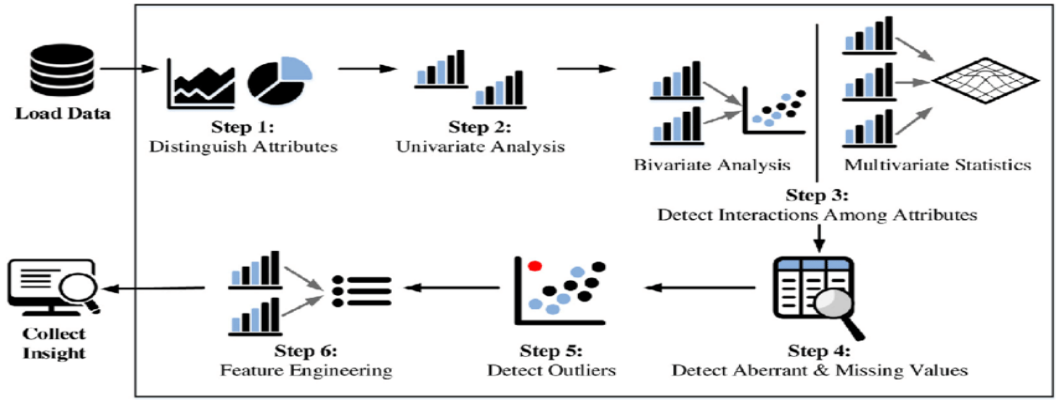

Figure 2: Key Techniques in Exploratory Data Analytics

## 3.1. Data Cleaning and Preprocessing

Data cleaning and preprocessing are crucial steps in exploratory data analysis (EDA) that involve identifying and handling issues in the dataset to ensure its quality and suitability for analysis. The main objectives of data cleaning and preprocessing are to address missing values, handle outliers, resolve inconsistencies, and transform the data into a usable format. Here are the key steps involved in data cleaning and preprocessing [6]:

**Handling Missing Values:** Missing values are problematic because they may introduce bias and affect the accuracy of the exploratory data analysis. Techniques for handling missing values, as indicated in Figure 3, include imputation, where missing values are filled in using statistical methods such as mean, median, or regression-based imputation. Alternatively, data points with missing values can be removed, but this should be done carefully to avoid significant data loss.

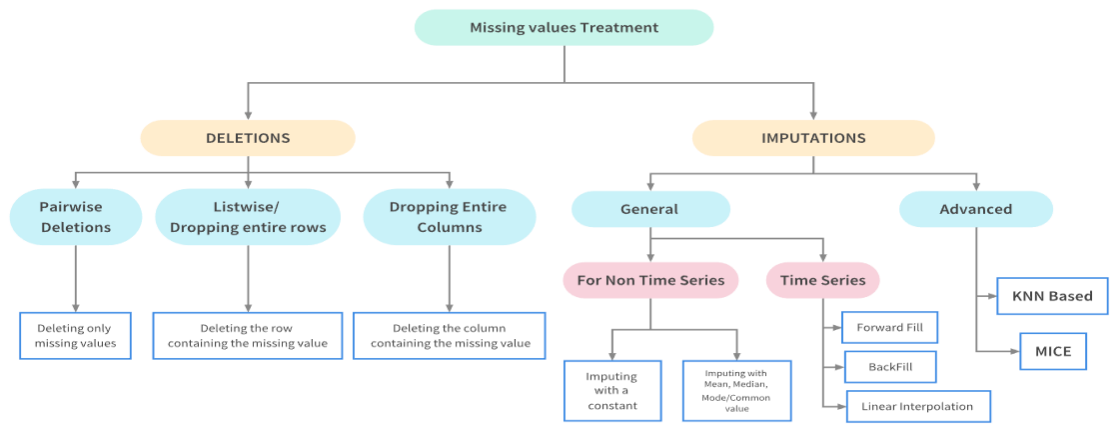

Figure 3: Techniques for handling missing values

**Outlier Detection and Handling:** Outliers are extreme or unusual data points that deviate significantly from the majority of the data. They can be caused by errors, data entry mistakes, or genuine anomalies. Outliers can impact the analysis, so it is important to identify and handle them appropriately. Outliers can be detected using statistical methods like z-score, IQR (interquartile range), or visualization techniques like box plots. Handling outliers can involve replacing them with a more reasonable value or removing them if they are deemed invalid [7].

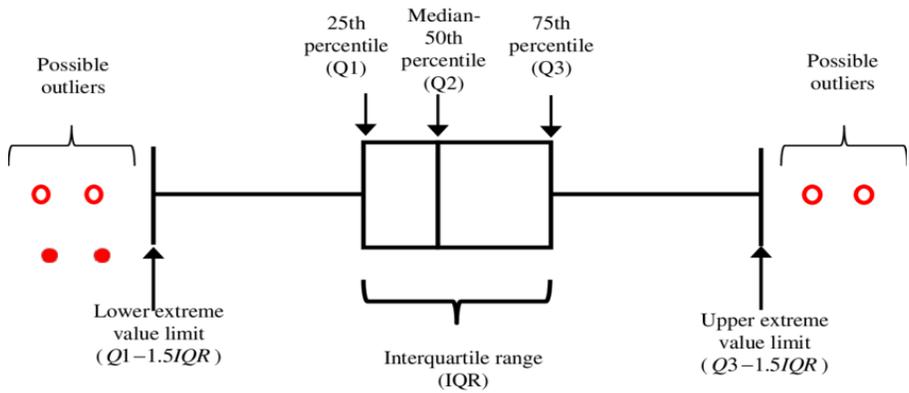

Figure 4: Outlier detection techniques

**Resolving Inconsistencies:** Inconsistencies in the data can arise from data entry errors, different data formats, or inconsistencies in data coding. It is important to identify and resolve these inconsistencies to ensure data accuracy and reliability. This may involve standardizing variables, correcting data entry errors, or converting data into a consistent format.

**Data Transformation:** Data transformation techniques are used to normalize the data or adjust its distribution. Common transformations include logarithmic transformation, square root transformation, or normalization to scale the data between a specific range. These transformations can help address issues such as skewed distributions or different scales among variables. Ensuring consistent formatting and encoding of data is also important for analysis. This includes converting variables into the appropriate data types (numeric, categorical, date/time) and encoding categorical variables using techniques like one-hot encoding or label encoding.

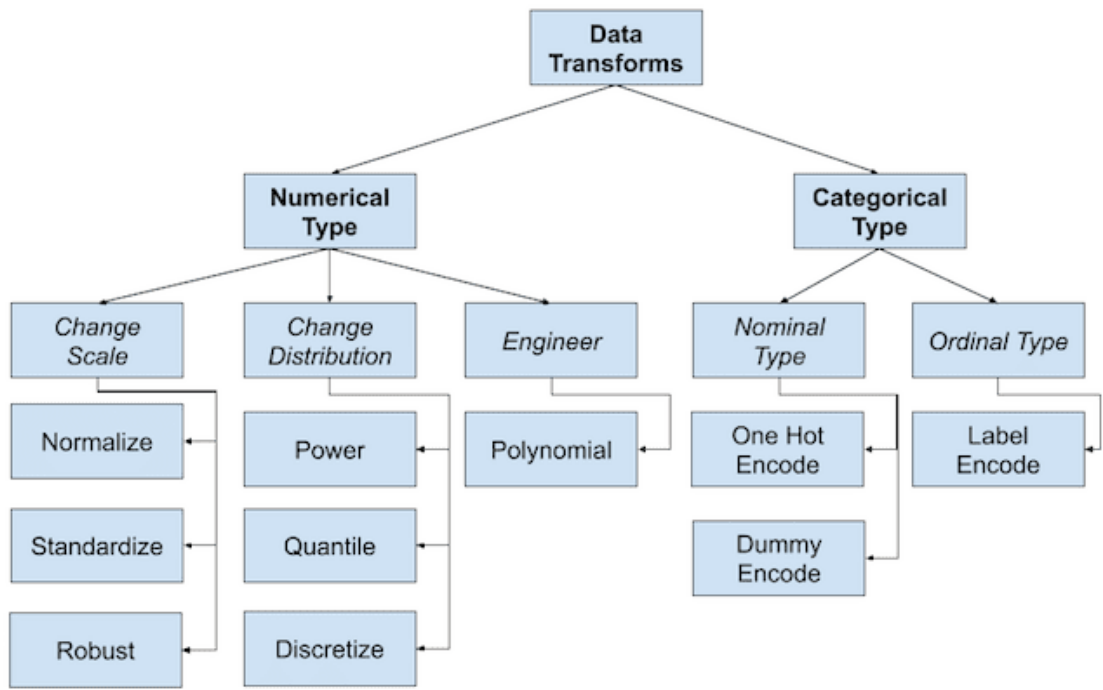

**Figure 5: Data Transforms techniques**

**Feature Engineering:** Feature engineering involves creating new features or transforming existing ones to extract relevant information and enhance the predictive power of the dataset. This may include creating interaction terms, polynomial features, or binning continuous variables into categorical variables [9].

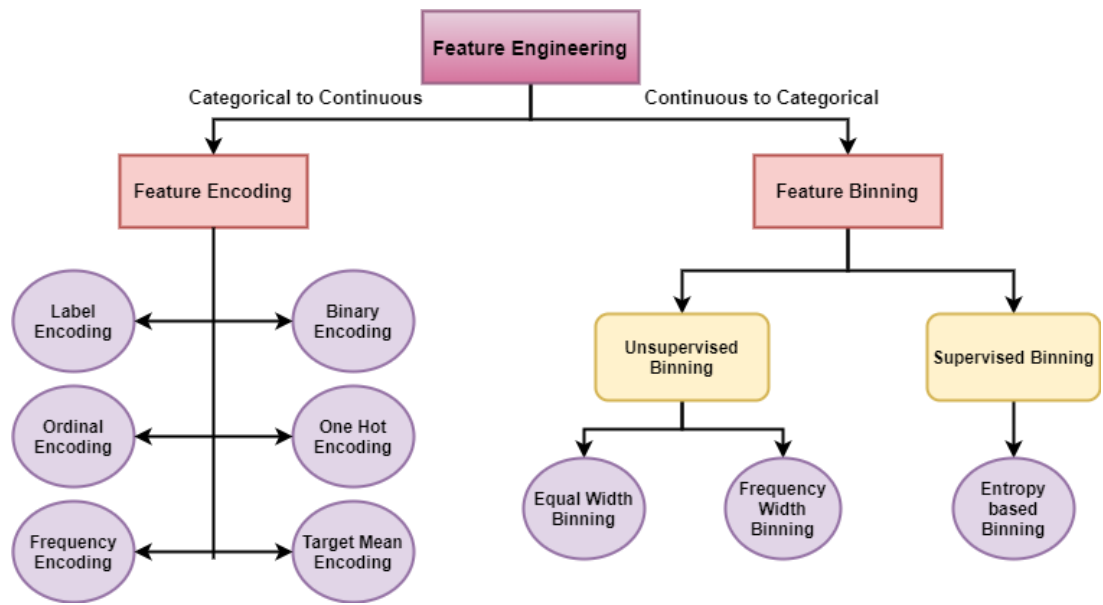

**Figure 6: Feature engineering techniques**

## 3.2. Summary Statistics

Summary statistics play a fundamental role in exploratory data analysis (EDA) by providing key insights into the central tendency, spread, and distribution of the data. They help in understanding the overall characteristics of the dataset without delving into complex statistical models. Here are some important summary statistics:

**Measures of Central Tendency:** These summary statistics measures the center or average of the data include the mean, median, and mode. The mean represents the arithmetic average of the data, the median is the middle value when the data is sorted, and the mode is the most frequently occurring value.

**Measures of Dispersion:** Summary statistics that quantify the spread or variability of the data include the range, variance, and standard deviation. The range is the difference between the maximum and minimum values in the dataset. The variance measures the average squared deviation from the mean, while the standard deviation is the square root of the variance, representing the average distance of data points from the mean [8].

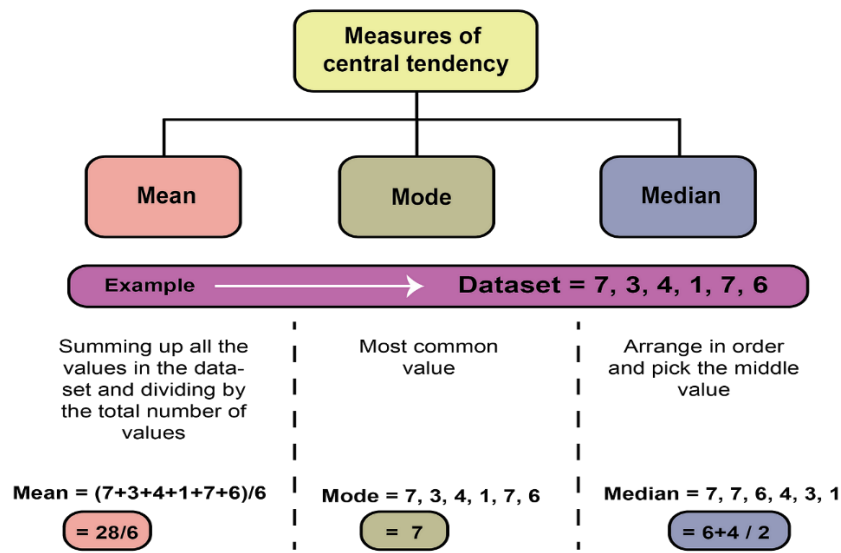

**Figure 7: Measure of central tendency**

**Percentiles:** Percentiles divide the data into equal portions, providing information on how values are distributed across the dataset. The median, for example, represents the 50th percentile, dividing the data into two equal halves. The quartiles divide the data into four equal parts: the first quartile (25th percentile), median (50th percentile), and third quartile (75th percentile).

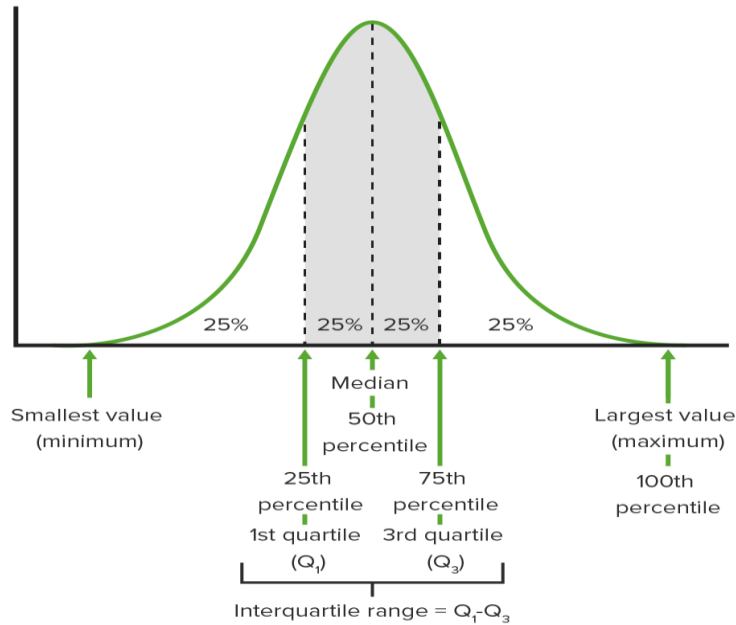

Figure 8: Percentiles description

**Skewness and Kurtosis:** Skewness measures the asymmetry of the data distribution, indicating whether it is skewed to the left (negative skew) or to the right (positive skew). Kurtosis measures the degree of peakedness or flatness of the data distribution, highlighting whether it has heavy tails or is more concentrated around the mean.

$$Skewness = \frac{3\,(Mean - Median)}{Std\ Deviation}$$

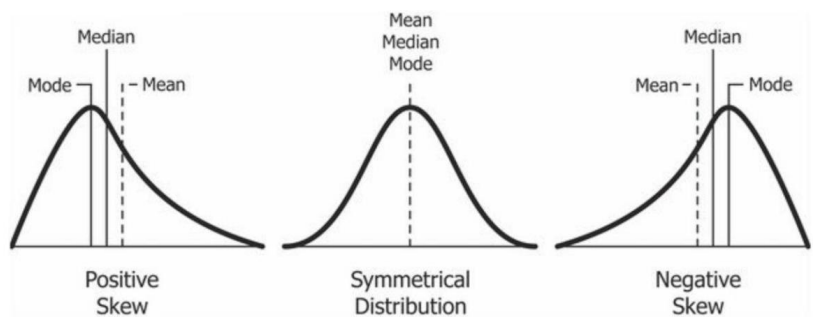

Figure 9: Types of Skewness

**Kurtosis** — Kurtosis describes the whether the data is light tailed (lack of outliers) or heavy tailed (outliers present) when compared to a Normal distribution. There are three kinds of Kurtosis:

- **Mesokurtic** — This is the case when the kurtosis is zero, similar to the normal distributions.
- **Leptokurtic** — This is when the tail of the distribution is heavy (outlier present) and kurtosis is higher than that of the normal distribution.
- **Platykurtic** — This is when the tail of the distribution is light (no outlier) and kurtosis is lesser than that of the normal distribution.

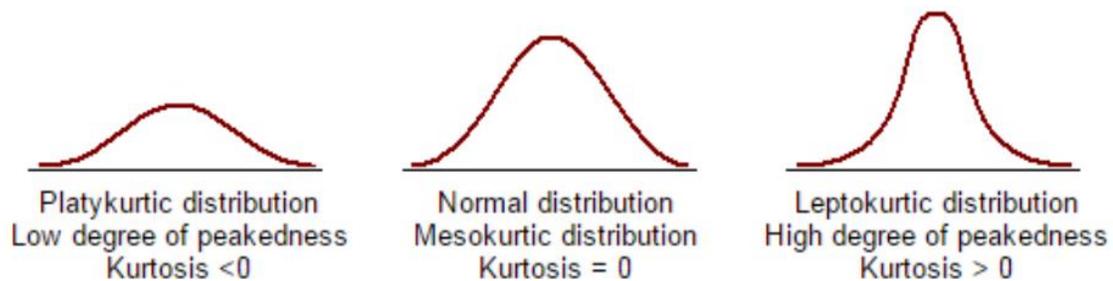

Figure 10: Types of Kurtosis

Summary statistics provide a snapshot of the data and help analysts gain an initial understanding of its characteristics. They can be used to identify outliers, assess the data's normality, and make comparisons between different datasets or subgroups. However, it's important to note that summary statistics alone do not provide a comprehensive analysis and should be used in conjunction with other exploratory techniques, visualizations, and hypothesis testing for a deeper understanding of the data.

## 3.3. Univariate, Bivariate and Multivariate Analysis

**Univariate Analysis:** Univariate analysis focuses on examining individual variables in isolation. It involves exploring the distribution, central tendency, dispersion, and shape of a single variable. Histograms, box plots, bar charts, and summary statistics are commonly used to visualize and analyze a single variable's characteristics [10].

**Distribution:** Distribution describes the pattern of values a variable takes on. It can be visualized using histograms, density plots, or kernel density estimations.

**Central Tendency:** Central Tendency measures the center or average of a variable's values. Common measures include the mean (average), median (middle value), and mode (most frequent value).

**Dispersion:** Dispersion describes the spread or variability of a variable's values. Measures such as variance and standard deviation quantify dispersion.

**Shape:** Shape represents the form or pattern of a variable's distribution. It can be characterized as symmetric, skewed (positively or negatively), or bimodal.

**Bivariate Analysis:** Bivariate analysis examines the relationship between two variables. It aims to understand how changes in one variable relate to changes in another variable. Scatter plots, line graphs, correlation analysis, and contingency tables are often used to explore associations, dependencies, and correlations between variables.

**Scatter Plot:** Scatter Plot is a graphical representation that displays the relationship between two continuous variables. It plots data points on a graph, where each point represents the values of the two variables.

**Correlation:** Correlation is a statistical measure that quantifies the strength and direction of the linear relationship between two continuous variables. It is typically represented by the correlation coefficient, such as Pearson's correlation coefficient.

**Contingency Table:** Contingency Table is also known as a cross-tabulation table. It displays the relationship between two categorical variables. It presents the counts or percentages of observations falling into different categories for each variable.

**Covariance:** Covariance measures the relationship between two variables. It indicates how changes in one variable are related to changes in another. Positive covariance indicates a direct relationship, while negative covariance indicates an inverse relationship.

**Multivariate Analysis:** Multivariate analysis explores the relationships between three or more variables simultaneously. It allows for a more comprehensive understanding of complex interactions and dependencies. Techniques such as principal component analysis (PCA), factor analysis, and multidimensional scaling are used to reduce dimensionality and visualize high-dimensional datasets.

## 3.4. Dimensionality Reduction

Dimensionality reduction is a technique used in exploratory data analysis (EDA) to reduce the number of variables or features in a dataset while preserving important information. High-dimensional datasets with numerous variables can pose challenges in analysis and interpretation. Dimensionality reduction aims to simplify the dataset by transforming it into a lower-dimensional space, making it more manageable and facilitating data visualization and modelling [11].

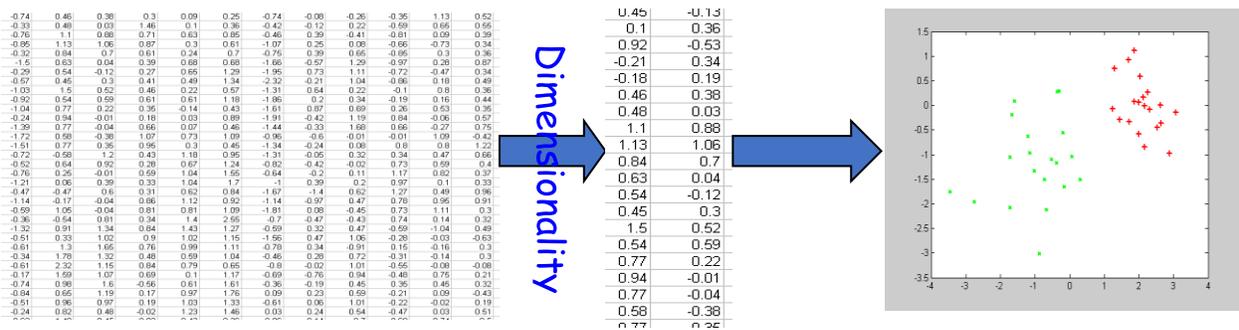

Figure 11: Dimensionality Reduction

The main objectives of dimensionality reduction are as follows:

**Curse of Dimensionality:** High-dimensional datasets can suffer from the curse of dimensionality, where the amount of data required to adequately cover the feature space grows exponentially with the number of dimensions. This can lead to sparse data, increased computational complexity, and difficulty in identifying meaningful patterns. Dimensionality reduction helps in mitigating these issues by reducing the dimensionality of the dataset.

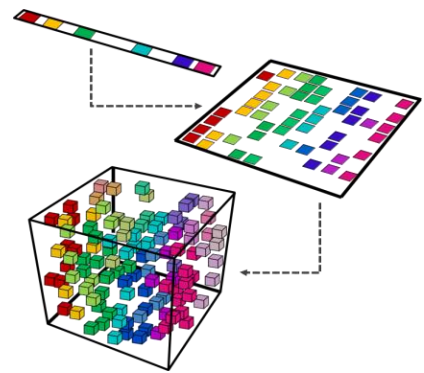

Figure 12: Curse of Dimensionality

**Dimensionality Reduction Overview:** Dimensionality reduction plays a crucial role in data analysis by selecting a subset of relevant features from the original dataset. This process identifies the most informative variables, it reduces the complexity of the analysis and improves model performance by focusing on the most discriminative features.

**Visualization:** Visualizing high-dimensional data is challenging. Dimensionality reduction techniques allow for the transformation of data into a lower-dimensional space (typically 2D or 3D), enabling the visualization of patterns, clusters, and relationships that may not be apparent in the original high-dimensional space. This facilitates data exploration and interpretation.

**Noise Reduction:** Dimensionality reduction can help in reducing noise or irrelevant information present in the dataset. By focusing on the most important features, it filters out noise and enhances the signal-to-noise ratio, leading to more accurate and robust analysis results.

**Approaches to Dimensionality Reduction**:

**Feature Selection:** This approach selects a subset of the original features based on certain criteria, such as statistical significance, importance, or correlation with the target variable. It aims to retain the most informative features while discarding redundant or irrelevant ones.

**Feature Extraction:** This approach transforms the original features into a new set of lower-dimensional features. Principal Component Analysis (PCA) is a widely used feature extraction technique that linearly transforms the data to orthogonal components, known as principal components, capturing the maximum amount of variance in the data.

Other popular dimensionality reduction techniques include t-Distributed Stochastic Neighbor Embedding (t-SNE), which emphasizes preserving the local structure of the data, and Linear Discriminant Analysis (LDA), which focuses on maximizing the class separability. Dimensionality reduction helps in simplifying complex datasets, improving visualization and interpretability, and enhancing the performance of subsequent analyses and models. It is a valuable tool in EDA to gain insights and extract meaningful information from high-dimensional data.

## 3.5. Data Visualization

Data visualization is a powerful technique used to explore and present data visually. Various visualization tools, such as histograms, scatter plots, box plots, and heatmaps, help analysts identify patterns, trends, and outliers more intuitively [12].

**Histogram:** A graphical representation of the distribution of a continuous variable. It displays the frequencies or relative frequencies of different intervals or bins.

**Box Plot:** Also known as a box-and-whisker plot, it provides a visual summary of the distribution of a continuous variable. It displays the median, quartiles, and potential outliers.

**Bar Chart:** A visual representation of categorical variables, where each category is represented by a bar. The height of the bar corresponds to the frequency or proportion of observations in that category.

**Heatmap:** A graphical representation that uses colors to represent the magnitude or intensity of a variable across different categories or time periods. It is particularly useful for displaying patterns in large datasets.

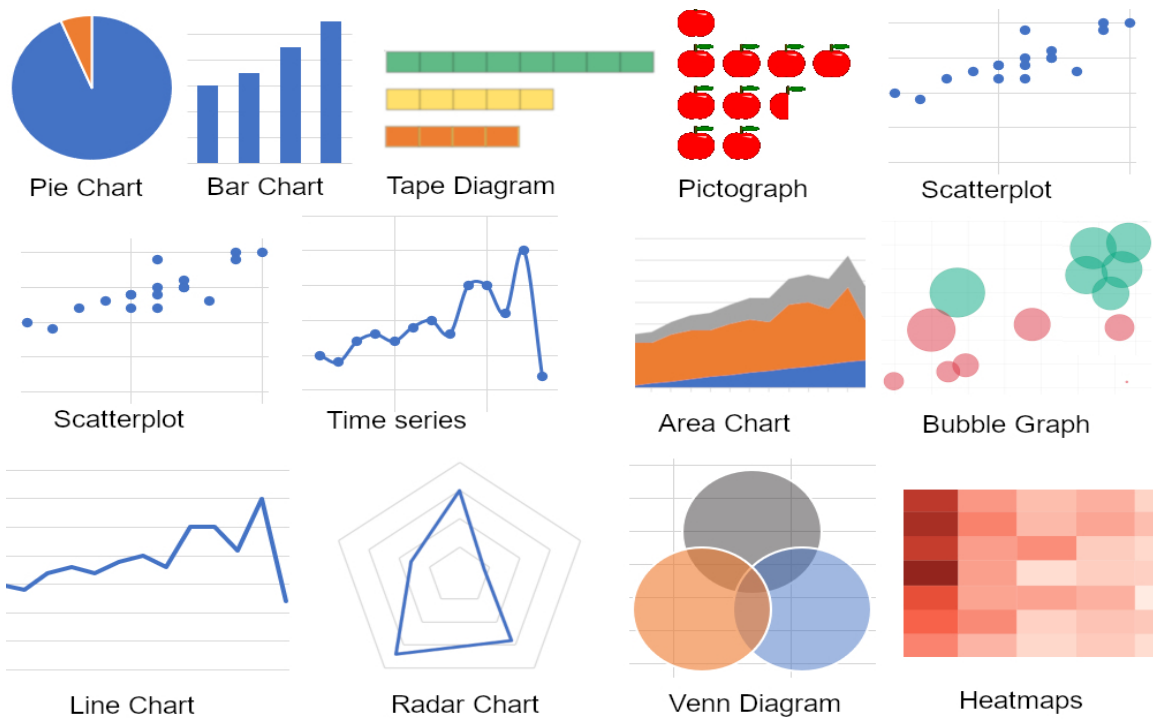

**Figure 13: Data Visualization**

## 3.6. Correlation Analysis

Correlation analysis techniques are used to assess the strength and direction of the relationship between two or more variables in a dataset. Correlation analysis helps in understanding the dependencies and associations between variables, which can provide insights into patterns, dependencies, and potential causal relationships. Here are some commonly used correlation analysis techniques [13]:

**Pearson's Correlation Coefficient:** Pearson's correlation coefficient measures the linear relationship between two continuous variables. It provides a value between -1 and +1, where -1 indicates a perfect negative correlation, +1 indicates a perfect positive correlation, and 0 indicates no correlation. The coefficient quantifies the strength and direction of the linear relationship between the variables.

**Spearman's Rank Correlation:** Spearman's rank correlation assesses the monotonic relationship between variables. It is used when the variables are measured on ordinal or non-linear scales. Spearman's correlation coefficient ranges between -1 and +1, with -1 indicating a perfect negative monotonic relationship, +1 indicating a perfect positive monotonic relationship, and 0 indicating no monotonic relationship.

**Kendall's Tau:** Kendall's tau is another non-parametric correlation measure that evaluates the strength and direction of the relationship between variables. It is particularly suitable for datasets with ties or when dealing with ordinal variables. Kendall's tau ranges between -1 and +1, with -1 indicating a perfect negative association, +1 indicating a perfect positive association, and 0 indicating no association.

**Point-Biserial Correlation:** The point-biserial correlation coefficient measures the correlation between a binary variable and a continuous variable. It determines the strength and direction of the relationship between the binary variable (coded as 0 and 1) and the continuous variable.

**Phi Coefficient:** The phi coefficient is used to assess the correlation between two binary variables. It is calculated by converting the data into a contingency table and applying the chi-square test to determine the association between the variables.

**Correlation Heatmap:** A correlation heatmap is a visual representation of the correlation matrix, where the magnitude of the correlation coefficients is represented by color. It provides a comprehensive overview of the correlations between multiple variables in a dataset, making it easier to identify patterns and relationships.

Correlation analysis techniques help in understanding the relationships between variables, identifying dependencies, and detecting potential multicollinearity issues in statistical modeling. However, it is important to note that correlation does not imply causation, and further analysis is required to establish causal relationships between variables.

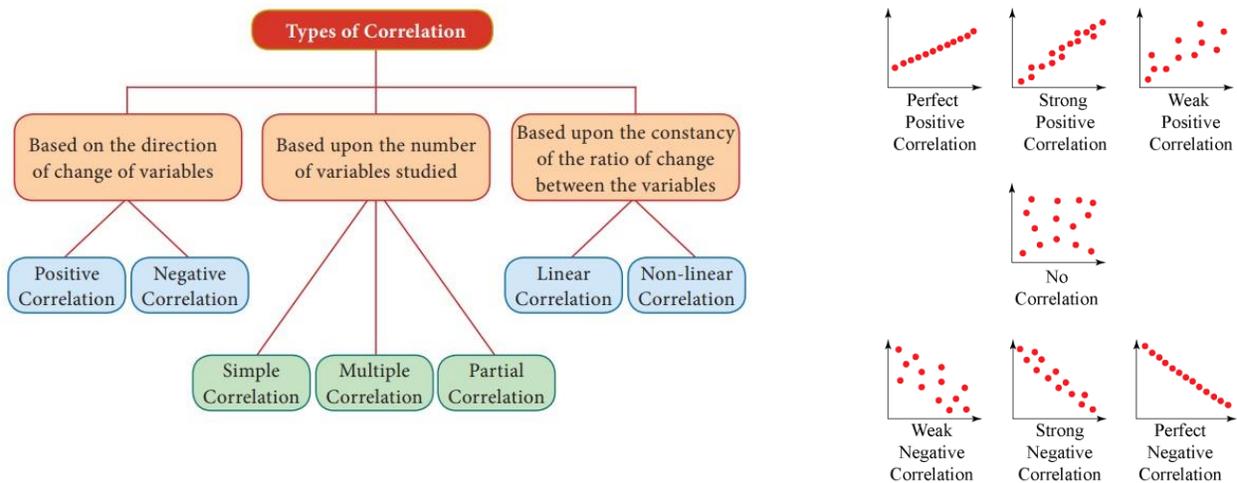

Figure 14: Correlation Analysis

## 3.7. Clustering Analysis

Clustering analysis techniques are used to group similar data points together based on their inherent characteristics or patterns. These techniques aim to discover hidden structures or clusters within the data, allowing for better understanding and organization of the dataset. Here are some commonly used clustering analysis techniques [14]:

**K-means Clustering:** K-means is one of the most widely used clustering algorithms. It partitions the data into a predefined number of clusters (k) by minimizing the within-cluster sum of squared distances. It iteratively assigns data points to the nearest centroid and updates the centroids until convergence. K-means is computationally efficient and effective for spherical clusters with similar sizes.

**Hierarchical Clustering:** Hierarchical clustering builds a hierarchy of clusters by either bottom-up (agglomerative) or top-down (divisive) approaches. Agglomerative clustering starts with each data point as a separate cluster and successively merges the closest clusters until a single cluster remains. Divisive clustering starts with the entire dataset as a single cluster and recursively divides it into smaller clusters. Hierarchical clustering produces a dendrogram, which provides insights into the clustering structure and allows for different levels of cluster granularity.

**Density-Based Spatial Clustering of Applications with Noise (DBSCAN):** DBSCAN is a density-based clustering algorithm that groups together data points within dense regions and identifies noise points as outliers. It defines clusters as areas of high density separated by areas of low density. DBSCAN requires specifying parameters such as minimum number of points and a distance threshold. It is particularly useful for detecting clusters of arbitrary shapes and handling outliers.

**Gaussian Mixture Models (GMM):** GMM is a probabilistic model that assumes the data is generated from a mixture of Gaussian distributions. It models the data as a combination of multiple Gaussian components, each representing a cluster. GMM assigns a probability of each data point belonging to each cluster and estimates the parameters (mean, covariance) of the Gaussian components through an iterative process. GMM is flexible in handling clusters of different shapes and can provide probabilistic cluster assignments.

**Self-Organizing Maps (SOM):** SOM, also known as Kohonen maps, is a neural network-based clustering technique. It uses unsupervised learning to map high-dimensional data onto a low-dimensional grid, where similar data points are grouped together. SOM involves an iterative training process where each data point is assigned to the closest node on the grid. It can preserve the topology and provide insights into the data distribution.

**Affinity Propagation:** Affinity Propagation is a clustering algorithm that does not require specifying the number of clusters in advance. It models data points as exemplars and iteratively exchanges messages between data points to determine which points should be exemplars. It is based on the concept of "message passing" and is suitable for datasets with no prior knowledge of the number of clusters.

Clustering analysis techniques help in identifying natural groupings, detecting outliers, and understanding the structure of the data. They are widely used in various domains, including customer segmentation, image recognition, social network analysis, and anomaly detection. The choice of clustering technique depends on the nature of the data, the desired output, and the specific problem at hand.

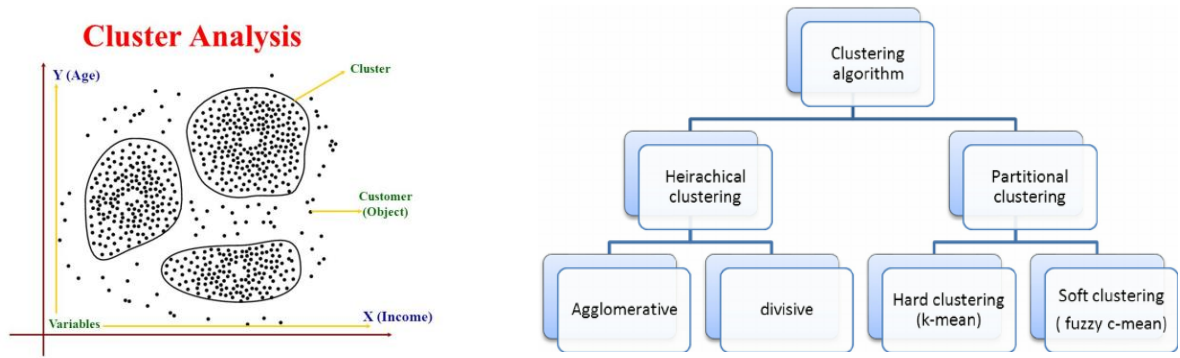

Figure 15: Clustering Analysis

## 3.8. Time Series analysis

Time series analysis techniques are used to analyze and forecast data that is collected over time. These techniques take into account the sequential nature of the data and help uncover patterns, trends, and dependencies that exist within the time series. Here are some commonly used time series analysis techniques [15]:

**Descriptive Statistics:** Descriptive statistics provide a summary of the basic properties of a time series, such as measures of central tendency (mean, median) and measures of dispersion (variance, standard deviation). They help in understanding the overall behavior and characteristics of the time series.

**Trend Analysis:** Trend analysis examines the long-term movement or direction of a time series. It involves identifying and modeling the underlying trend component, which represents the overall pattern in the data. Trend analysis techniques include simple moving averages, exponential smoothing, and regression analysis.

**Seasonal Analysis:** Seasonal analysis focuses on identifying and modeling repeating patterns or seasonal effects within a time series. It helps in understanding the regular fluctuations that occur at fixed intervals (e.g., daily, monthly, yearly). Seasonal analysis techniques include seasonal decomposition of time series, seasonal indices, and seasonal autoregressive integrated moving average (SARIMA) models.

**Autocorrelation Analysis:** Autocorrelation analysis examines the correlation between a time series and its lagged values. It helps in identifying dependencies and patterns that exist within the series. Autocorrelation function (ACF) and partial autocorrelation function (PACF) plots are commonly used to visualize and interpret the autocorrelation structure of a time series.

**Stationarity Analysis:** Stationarity is an important concept in time series analysis. It refers to the property of a time series where the statistical properties (mean, variance) remain constant over time. Stationarity analysis involves testing and transforming the time series to achieve stationarity, as non-stationarity can affect the accuracy of modeling and forecasting. Techniques such as differencing and transformation are commonly used to achieve stationarity.

**Time Series Modeling:** Time series modeling involves constructing mathematical models to represent the underlying structure and dynamics of a time series. Popular time series models include autoregressive integrated moving average (ARIMA), seasonal ARIMA (SARIMA), exponential smoothing models, and state space models. These models capture the dependencies and patterns observed in the data and can be used for forecasting future values.

**Time Series Forecasting:** Time series forecasting predicts future values of a time series based on historical data and the identified patterns and trends. Various techniques, such as ARIMA, SARIMA, exponential smoothing, and machine learning algorithms, can be used for time series forecasting. The choice of the forecasting method depends on the characteristics of the data and the desired level of accuracy.

Time series analysis techniques are widely used in finance, economics, sales forecasting, weather forecasting, and many other domains. They provide valuable insights into the behavior and dynamics of time-dependent data and support decision-making processes by predicting future values and understanding the factors that influence the time series.

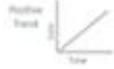

**Figure 16: Time Series Analysis**

## 4. Application of EDA in Banking and Finance

Exploratory Data Analysis (EDA) plays a crucial role in the banking and finance industry. Here are some key applications of EDA in this domain:

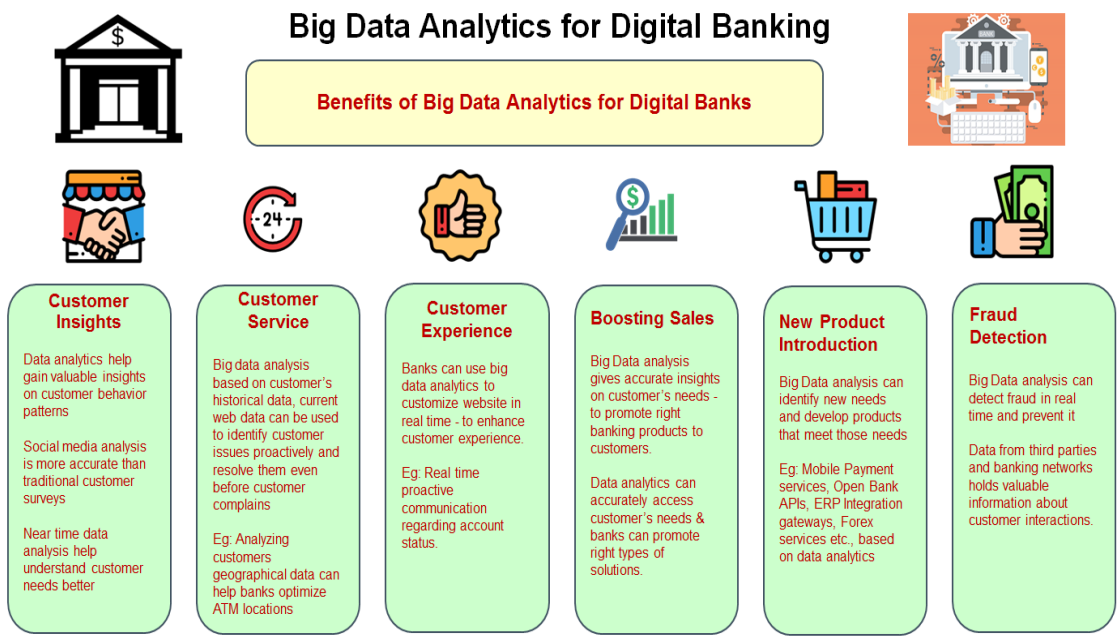

**Figure 17: Application of EDA in Banking and Finance**

**Fraud Detection:** EDA helps in identifying patterns and anomalies in financial transactions that may indicate fraudulent activities. By analyzing transactional data, EDA techniques can uncover suspicious patterns, such as unusual transaction amounts, frequent transfers, or irregular spending patterns, which can aid in fraud detection and prevention [16].

**Customer Segmentation:** EDA allows banks and financial institutions to segment their customer base by analyzing various demographic, transactional, and behavioral data. By identifying distinct customer segments, banks can tailor their marketing strategies, product offerings, and customer service to meet the specific needs of different customer groups [17].

**Credit Risk Assessment:** EDA is used to analyze credit-related data, such as credit scores, payment history, and financial information, to assess the creditworthiness of individuals and businesses. By identifying patterns and trends in historical credit data, EDA helps in predicting and evaluating the risk associated with lending money or extending credit [18].

**Loan Portfolio Analysis:** EDA analyzes the performance and characteristics of loan portfolios. By examining historical loan data, EDA techniques can provide insights into factors such as default rates, loan repayment patterns, and risk exposure. This information enables banks to make informed decisions regarding loan origination, loan pricing, and portfolio management [19].

**Customer Churn Analysis:** EDA is useful in understanding customer behavior and identifying factors that contribute to customer churn. By analyzing customer data, transactional patterns, and demographic information, EDA techniques can uncover insights into customer attrition. This knowledge allows banks to develop targeted retention strategies and improve customer satisfaction [20].

**Market Analysis:** EDA is utilized in analyzing market trends, investment patterns, and financial market data. By examining historical market data and conducting exploratory analysis, banks can identify patterns, correlations, and anomalies that can inform investment strategies, risk management decisions, and portfolio diversification [21].

**Compliance and Risk Management:** EDA is employed to analyze regulatory compliance data, transactional data, and risk-related information. By identifying patterns and anomalies, EDA helps banks ensure compliance with regulations, detect potential fraud or money laundering activities, and manage operational and financial risks effectively [22].

Overall, EDA enables data-driven decision-making in the banking and finance industry, facilitating risk assessment, fraud detection, customer segmentation, and market analysis. It helps banks gain insights from their vast amounts of data, leading to improved operational efficiency, enhanced risk management, and better customer experiences.

## 5. EDA on churning of customers in Banking Sector

In the banking sector, customer churn refers to the phenomenon where customers discontinue their relationship with a bank or stop using a particular product or service offered by the bank, such as credit card services. Customer churn poses a significant challenge for banks as it can lead to a loss of revenue, market share, and customer loyalty. Customer churn in the banking sector can occur due to various reasons, including dissatisfaction with service quality, high fees, better offers from competitors, changing financial needs, or a negative customer experience. Identifying and predicting customer churn is crucial for banks to implement proactive retention strategies and minimize customer attrition.

By analyzing customer data and employing techniques such as Exploratory Data Analysis (EDA), banks can gain insights into the factors influencing customer churn. EDA allows banks to explore the data, identify patterns, and uncover key drivers that contribute to customer attrition. This analysis can involve examining customer demographics, transactional behavior, account activity, customer feedback, and other relevant data points.

Once the potential churn factors are identified, banks can develop targeted retention initiatives to retain at-risk customers. This may involve personalized communication, tailored offers, improved customer service, loyalty programs, or incentives to encourage customers to stay with the bank.

## 5.1. Dataset Description

To perform operations, exploratory data analysis (EDA), visualization, and analysis on the Churn Modelling dataset from Kaggle, we will utilize various libraries in Python. Firstly, we need to import necessary libraries such as Pandas and Matplotlib. Pandas provides powerful data manipulation and analysis tools, while Matplotlib aids in creating visualizations to gain insights from the data. By using these libraries, we can extract meaningful information, identify patterns, and understand customer churn in the banking industry.

The dataset used in this project comprises comprehensive information about the bank customers. The dataset includes various attributes related to each customer, with the target variable is represented as a binary value indicating whether the customer has closed their bank account or remains an active customer. This target variable is a crucial for analyzing customer churn within the banking sector.

The dataset is taken from Kaggle repository [23]. There are 14 attributes in the dataset. The information of the dataset is shown as follows in figure:

**RowNumber** — the record (row) number and has no effect on the output.
**CustomerId** — contains random values and has no effect on customer leaving the bank.
**Surname** — the surname of a customer has no impact on their decision to leave the bank.
**CreditScore** — can have an effect on customer churn, since a customer with a higher credit score is less likely to leave the bank.
**Geography** — a customer's location can affect their decision to leave the bank.
**Gender** — it's interesting to explore whether gender plays a role in a customer leaving the bank. We'll include this column, too.
**Age** — this is certainly relevant, since older customers are less likely to leave their bank than younger ones.
**Tenure** — refers to the number of years that the customer has been a client of the bank. Normally, older clients are more loyal and less likely to leave a bank.
**Balance** — also a very good indicator of customer churn, as people with a higher balance in their accounts are less likely to leave the bank compared to those with lower balances.
**NumOfProducts** — refers to the number of products that a customer has purchased through the bank.
**HasCrCard** — denotes whether or not a customer has a credit card. This column is also relevant, since people with a credit card are less likely to leave the bank. (0=No,1=Yes)
**IsActiveMember** — active customers are less likely to leave the bank, so we'll keep this. (0=No,1=Yes)
**EstimatedSalary** — as with balance, people with lower salaries are more likely to leave the bank compared to those with higher salaries.
**Exited** — whether or not the customer left the bank. This is what we have to predict. (0=No,1=Yes)

**Figure 18: Dataset Description**

## 5.2. EDA steps on Bank Churning dataset

Our main aim is to anticipate whether clients will opt to terminate their banking relationship in the near future. We have access to historical data that encompasses customer behavior patterns and instances of contract terminations with the bank. Several key analyses are required:

- Determine the target feature and the labels in target.
- Analyze the correlation between target and discrete/continuous features.
- Analyze the correlation among different features.
- Determine the number of missing values.
- Analyze and determine the possible outlier data.
- Plan to use what strategies to handle missing values and outliers.

Following are the details of the steps taken:

**STEP 1:** Importing required libraries

```python
import pandas as pd
import numpy as np
import matplotlib.pyplot as plt
import seaborn as sns
import warnings

sns.set()
warnings.simplefilter('ignore')
```

**Getting the data**

```python
data = pd.read_csv('Churn_Modelling.csv')
df = data.copy()
df.head()
```

The data contains 10000 rows and 14 columns.

**STEP 2:** Perform Data wrangling to convert raw data into a usable form such as removing the unimportant field and understanding the missing values, null values and taking appropriate measures.

```
df.drop(['RowNumber', 'CustomerId', 'Surname'], axis=1, inplace=True)
df.head()
```

|   | CreditScore | Geography | Gender | Age | Tenure | Balance | NumOfProducts | HasCrCard | IsActiveMember | EstimatedSalary | Exited |
|---|---|---|---|---|---|---|---|---|---|---|---|
| 0 | 619 | France | Female | 42 | 2 | 0.00 | 1 | 1 | 1 | 101348.88 | 1 |
| 1 | 608 | Spain | Female | 41 | 1 | 83807.86 | 1 | 0 | 1 | 112542.58 | 0 |
| 2 | 502 | France | Female | 42 | 8 | 159660.80 | 3 | 1 | 0 | 113931.57 | 1 |
| 3 | 699 | France | Female | 39 | 1 | 0.00 | 2 | 0 | 0 | 93826.63 | 0 |
| 4 | 850 | Spain | Female | 43 | 2 | 125510.82 | 1 | 1 | 1 | 79084.10 | 0 |

```
df.isnull().sum().to_frame('No. of Nulls')
```

Once data is clean and not null. We may perform the Exploratory Data Analysis.

**STEP 3:** Exploratory Data Analysis

In this dataset, we aim to understand whether a customer will continue banking with us or decide to leave. From the statistics provided above, we observe that the majority of customers are from France, and a significant portion of them are male.

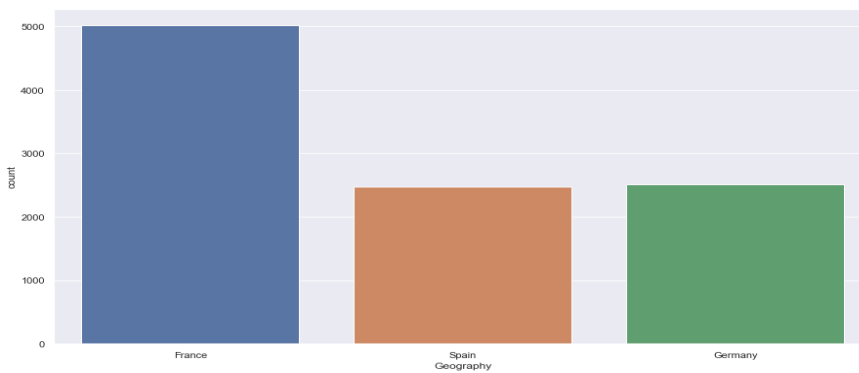

**Figure 19: Graph between Customer count and geography**

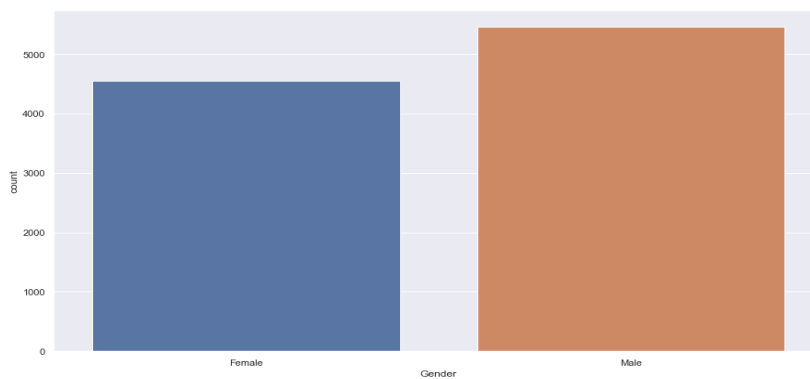

**Figure 20: Graph between Customer count and Gender**

We have also plotted the heatmap to get information about the correlation among different columns.

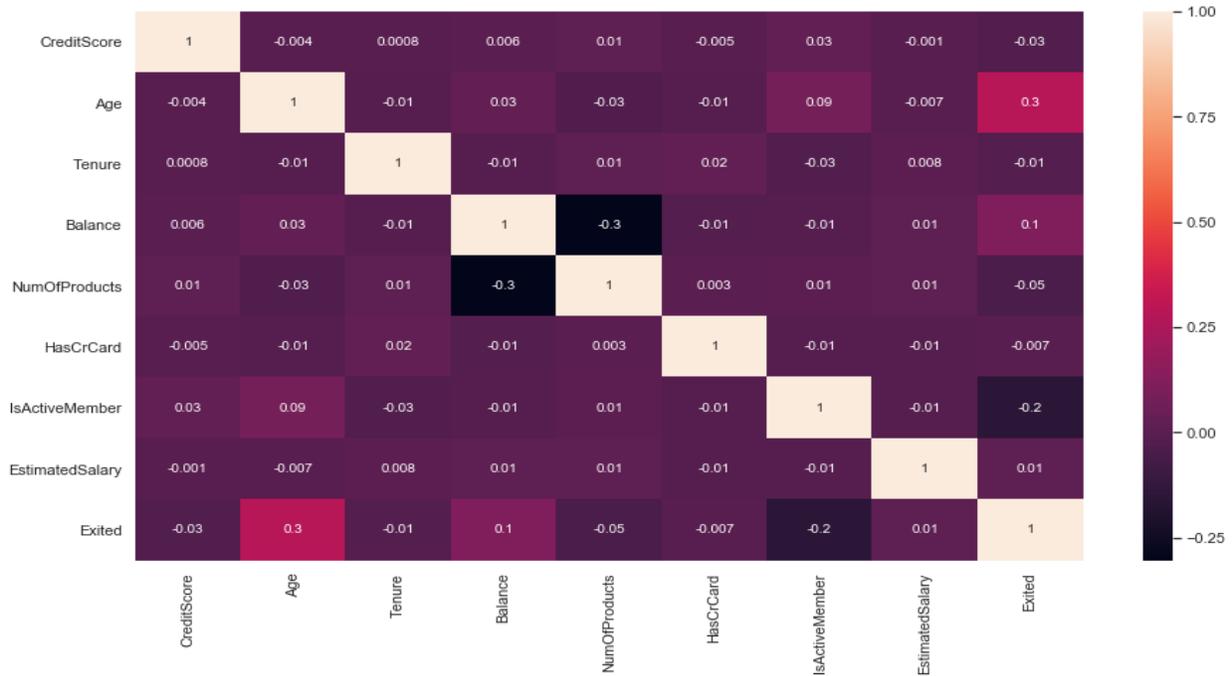

**Figure 21: Heatmap**

We have also plotted the distribution of Credit Score. Here, most of the distributions are between 600 and 700 and the distribution is normal (i.e. not skewed).

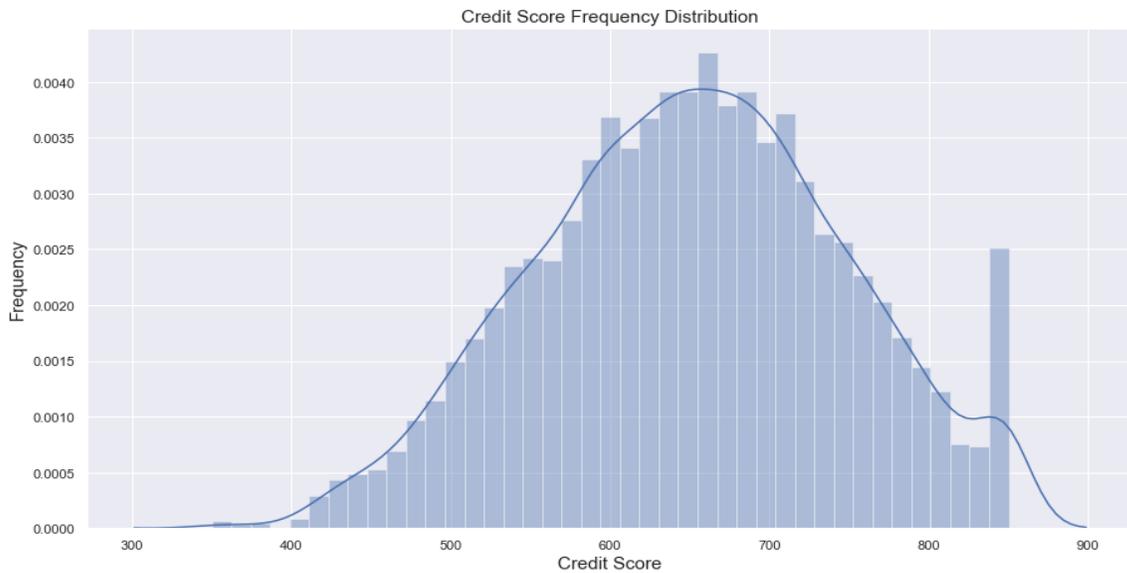

**Figure 22: Distribution of Credit Score**

After plotting the credit score with age. It is evident that there is no correlation among age and credit score.

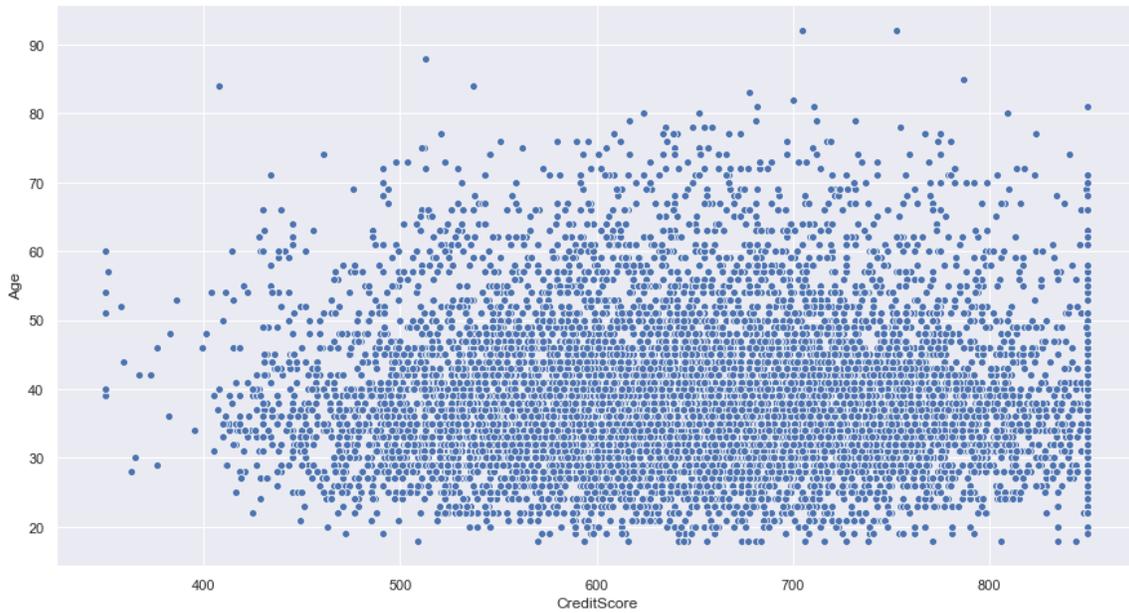

**Figure 23: Graph between Credit Score and Age**

The customer tenure with the bank was also studied and it is found that the people have similar years of tenure. However, there are a few people with less than a year or 10 years.

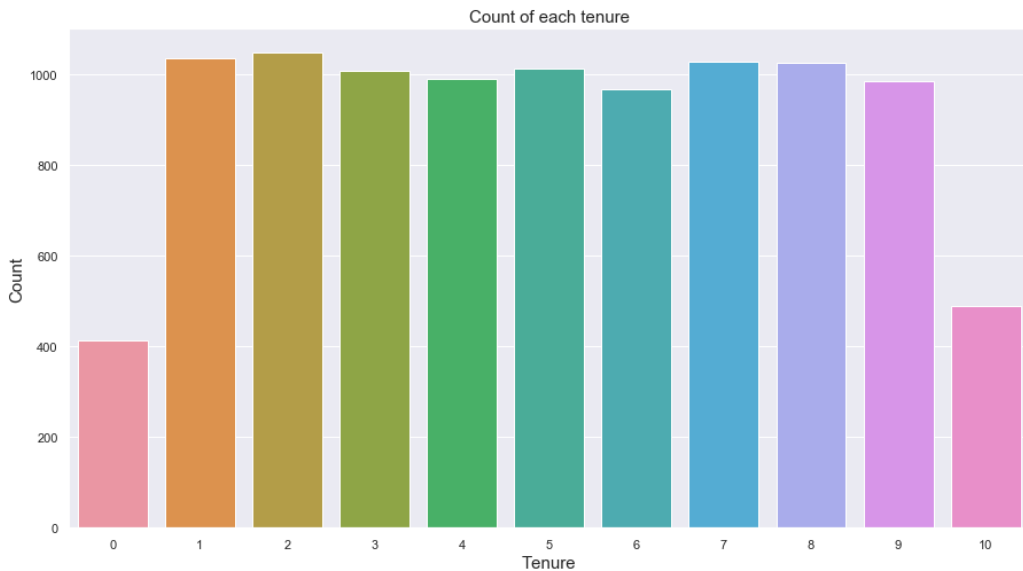

**Figure 24: Graph between Customer Count and Tenure**

The graph below represents the relationship of Geography with Churn/Exited. It is evident that all the counties are having similar pattern of exiting.

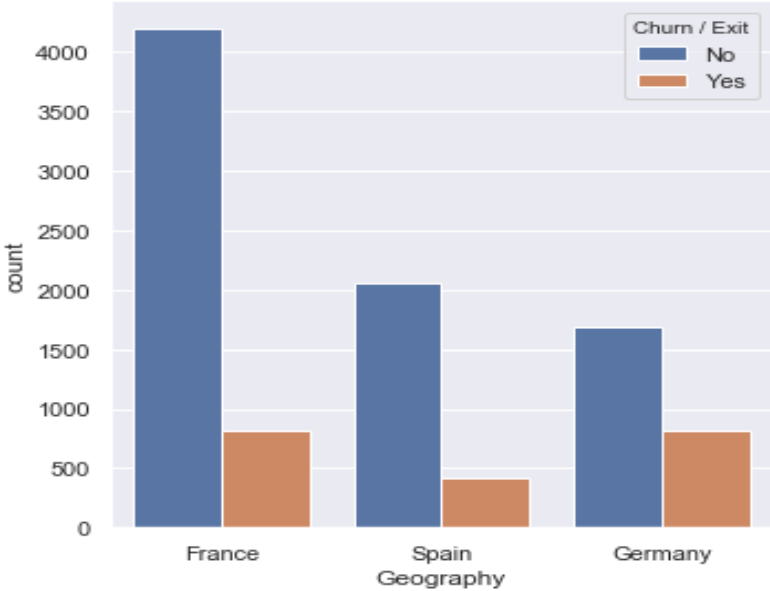

**Figure 25: Relationship of Geography with Churn / Exited**

In conclusion, a comprehensive analysis of the dataset has revealed several key observations. The 'CreditScore' feature exhibits a distribution that is close to normal, with an exceptional outlier at the maximum value of 850, suggesting the presence of the highest possible credit rating within the bank. The 'Age' feature follows a normal distribution with a right skew, while notable outliers are observed at values that are multiples of 10, specifically 30, 40, 50, 60, and 70. Additionally, the 'Tenure' feature demonstrates a distribution that is approximately uniform. The 'Balance' feature follows a normal distribution, albeit with a significant outlier at zero, indicating a considerable number of customers who do not maintain funds in their accounts. Moreover, the majority of customers in the dataset possess 1 or 2 products, while the 'EstimatedSalary' feature exhibits a uniform distribution. It is noteworthy that the bank has an almost equal number of male and female clients, as well as an equitable distribution between active and inactive clients. Interestingly, 71% of the customers hold a bank credit card, while approximately 20% of the customers have discontinued their usage of the bank's services. These insights provide a valuable understanding of the dataset, enabling further exploration and the potential development of targeted strategies for customer retention and satisfaction.

# 6. Challenges and considerations

Performing Exploratory Data Analysis (EDA) in the banking and finance industry comes with its own set of challenges and considerations [24]. Here are some key factors to keep in mind:

**Data Quality and Availability:** One of the primary challenges in EDA is ensuring the quality and availability of data. In banking and finance, data sources may be vast and complex, with different data formats and data quality issues. Missing data, inconsistencies, and errors can impact the accuracy and reliability of the analysis. Employing data cleansing and preprocessing techniques is necessary to address these challenges.

**Data Privacy and Security:** Banks manages sensitive customer information, financial transactions, and compliance-related data. Maintaining data privacy and security throughout the EDA process is crucial. Compliance with regulations, such as the General Data Protection Regulation (GDPR), and ensuring data anonymization or pseudonymization are essential to protect customer privacy.

**Data Integration:** Data in Banking and finance often reside in various systems and databases. Integrating data from multiple sources and ensuring data consistency and compatibility can be challenging. Proper data integration techniques are needed to create a unified and comprehensive dataset for analysis.

**Dimensionality and Scale:** Banking and finance datasets can be voluminous and high-dimensional, containing numerous variables and attributes. Handling large datasets and dealing with high dimensionality pose challenges in terms of computational complexity and visualization. Dimensionality reduction techniques, such as feature selection or extraction, may be necessary to reduce complexity and improve analysis efficiency.

**Complex Relationships and Dependencies:** Financial data often exhibits complex relationships and dependencies that simple statistical techniques may not capture. Non-linear patterns, time-varying relationships, and hidden dependencies can pose challenges in uncovering meaningful insights. Advanced analytics techniques, such as machine learning algorithms, may be required to address these complexities.

**Interpretability and Explainability:** It is crucial for the banking and finance industry to have transparent and interpretable analysis results. Stakeholders need to understand the rationale behind the findings and decisions based on EDA. Ensuring the interpretability and explainability of models and analysis outputs is essential for gaining trust and making informed business decisions.

**Regulatory and Compliance Considerations:** The banking and finance industry is heavily regulated, and compliance with regulatory requirements is paramount. EDA should adhere to regulatory guidelines and consider the legal and ethical implications of the analysis. Compliance with data protection laws, anti-money laundering regulations, and risk management frameworks should be taken into account.

**Domain Expertise:** EDA in banking and finance requires a deep understanding of the industry, financial products, risk management, and regulatory frameworks. Domain expertise is essential to ask relevant questions, interpret the results, and derive actionable insights from the analysis.

By addressing these challenges and considerations, EDA in banking and finance can provide valuable insights, support decision-making processes, and help in risk management, compliance, customer satisfaction, and overall business performance.

## 7. Conclusion

Exploratory Data Analytics (EDA) serves as a vital tool for understanding and gaining insights from complex datasets. By employing various techniques such as data cleaning, descriptive statistics, data visualization, correlation analysis, dimensionality reduction, and clustering analysis, analysts can uncover patterns, relationships, and anomalies within the data. EDA finds application in diverse domains, including business analytics, healthcare, finance, social sciences, manufacturing, and environmental studies. However, analysts must be aware of challenges such as data quality, bias, confounding factors, overfitting, and ethical considerations throughout the EDA process. By leveraging EDA effectively, organizations and researchers can make data-informed decisions, drive innovation, and unlock hidden insights that can contribute to better outcomes in various fields.